\documentclass[aps,nofootinbib,superscriptaddress
]{revtex4-2}
\usepackage{graphicx}
\usepackage{amsmath}
\usepackage{amsfonts}
\usepackage{amssymb}
\usepackage{color}

\newcommand{\hide}[1]{}

\usepackage{color}
\def\red{\color{red}}

\begin{document}

\title{Catastrophic Emission of Charges from Near-Extremal Charged Nariai Black Holes. II. Rotation Effect}

\author{Chiang-Mei Chen} \email{cmchen@phy.ncu.edu.tw}
\affiliation{Department of Physics, National Central University, Chungli 32001, Taiwan}
\affiliation{Center for High Energy and High Field Physics (CHiP), National Central University, Chungli 32001, Taiwan}

\author{Chun-Chih Huang} \email{makedate0809@gmail.com}
\affiliation{Department of Physics, National Central University, Chungli 32001, Taiwan}

\author{Sang Pyo Kim} \email{sangkim@kunsan.ac.kr}
\affiliation{Department of Physics, Kunsan National University, Kunsan 54150, Korea}
\affiliation{Asia Pacific Center for Theoretical Physics, Pohang 37673, Korea}
\affiliation{Center for High Energy and High Field Physics (CHiP), National Central University, Chungli 32001, Taiwan}

\author{Chun-Yu Wei} \email{weijuneyu@gmail.com}
\affiliation{Department of Physics, National Central University, Chungli 32001, Taiwan}

\date{\today}

\begin{abstract}
Kerr-Newman black holes in a de Sitter (dS) space have the limit of rotating Nariai black holes with the near-horizon geometry of a warped ${\rm dS}_3 \times {\rm S}^1/Z_2$ when the black hole horizon and the cosmological horizon coincide or approach close to each other. We study the rotation effect on the spontaneous emission of charges in the near-extremal rotating charged Nariai black hole and compare it to those from the near-extremal Nariai black hole in Phys. Rev. D \textbf{110}, 085020 (2024) and near-extremal Kerr-Newman black hole in de Sitter space in Eur. Phys. J. C \textbf{83}, 219 (2023). In strong contrast to the near-extremal Kerr-Newman black hole in dS space, the near-extremal rotating Nariai black hole also has an exponential amplification for the emission of high energy charges, which becomes catastrophic regardless of angular momentum when two horizons coincide. The radius of rotating Nariai black holes monotonically increases as the angular momentum and charge of black holes increase, which gives a weaker electric field on the horizon than Nariai black holes. Thus the angular momentum of black holes that drags particles on the horizon decreases the mean number of charges by a factor not by an order. We observe a catastrophic emission of boson condensation for charges with an effective energy equal to the chemical potential in the spacelike outer region of the cosmological horizon. Remarkably, the Schwinger emission of charges in the standard particle model may prevent the rotating Nariai black holes from evolving into spacetimes with a naked singularity when the angular momentum is close to the allowed maximum, which Nariai black holes cannot avoid.
\end{abstract}


\maketitle

\section{Introduction}
Extremal or near-extremal charged black holes in which the inner and outer horizons coincide or close to each other can emit charges via the Schwinger pair production near the horizon though the Hawking radiation is exponentially suppressed due to vanishingly small Hawking temperature. Furthermore, the enhanced symmetry of near-horizon geometry of near-extremal black holes~\cite{Bardeen:1999px} allows one to explicitly find the solutions of the Klein-Gordon or Dirac equation for charged scalars or fermions, and therefrom the mean number or the vacuum persistence amplitude and the absorption rate from the near-extremal black holes have been found in asymptotically flat space~\cite{Chen:2012zn, Chen:2014yfa, Chen:2016caa, Chen:2017mnm} and in the (anti-)de Sitter space~\cite{Chen:2020mqs, Chen:2021jwy}. The near-horizon geometry of ${\rm AdS}_2 \times {\rm S}^2$ in non-rotating charged black holes and the warped ${\rm AdS}_3 \times {\rm S}^1/Z_2$ in rotating charged black holes lead to a universal formula~\cite{Chen:2012zn, Chen:2014yfa, Chen:2016caa, Chen:2017mnm, Chen:2020mqs, Chen:2021jwy} for the charge emission in terms of the effective temperature for charges accelerated by the electric field on the horizon and the two-dimensional curvature of ${\rm AdS}_2$, which can be explained by the Schwinger pair production by a constant electric field in ${\rm AdS}_2$~\cite{Cai:2014qba}. Moreover, the emissions that can cause the black holes evolving into singular spacetimes are in fact forbidden by the Breitenlohner-Friedman (BF) bound.

Charged black holes in de Sitter (dS) space have another horizon, the cosmological horizon, in which Nariai black holes are obtained by coinciding the black hole horizon with the cosmological horizon~\cite{Romans:1991nq}. Using the enhanced geometry of ${\rm dS}_2 \times {\rm S}^2$ for (near-)extremal Nariai black holes, the spontaneous emission of charges via the Schwinger pair production has been studied~\cite{Chen:2023swn}. Remarkably, the effective temperature for charge emission is determined by the Unruh temperature for accelerating charges on the horizon and the Gibbons-Hawking temperature whose square is proportional to the two-dimensional curvature of ${\rm dS}_2$~\cite{Cai:2014qba}. The geometry of ${\rm dS}_2$ for Nariai black holes enhances the spontaneous emission of charges due to the Gibbons-Hawking radiation from the cosmological horizon: the black hole horizon emits charges of the same kind of and the cosmological horizon emits charges of the opposite kind of the black hole charge. However, the charged Nariai black holes do not have the BF bound that guarantees the stability of charges in an electric field in anti-de Sitter (AdS) space~\cite{Breitenlohner:1982jf}. In contrast, spontaneous emissions driving the charged Nariai black holes into singular spacetimes are physically allowed. This means that the cosmic censorship can be violated~\cite{Montero:2019ekk, Chen:2023swn}.

The main purpose of this paper is to study, in the Einstein-Maxwell theory with a positive cosmological constant $\Lambda$, the spontaneous emission of charges from rotating charged Nariai black holes that are the coincidence limit of the black hole horizon and the cosmological horizon. Particularly we will focus the study on the rotation effect on the charge emission from near-extremal rotating charged Nariai black holes and discuss the physical implications in various limits: the (non-rotating) charged Nariai black holes, rotating uncharged Nariai black holes and nonrotating uncharged Nariai black holes. We will also compare the emission of charges from near-extremal rotating charged Nariai black holes with those from near-extremal Nariai black holes and near-extremal KN-dS black holes.

\begin{figure}
\includegraphics[scale=0.2, angle=0]{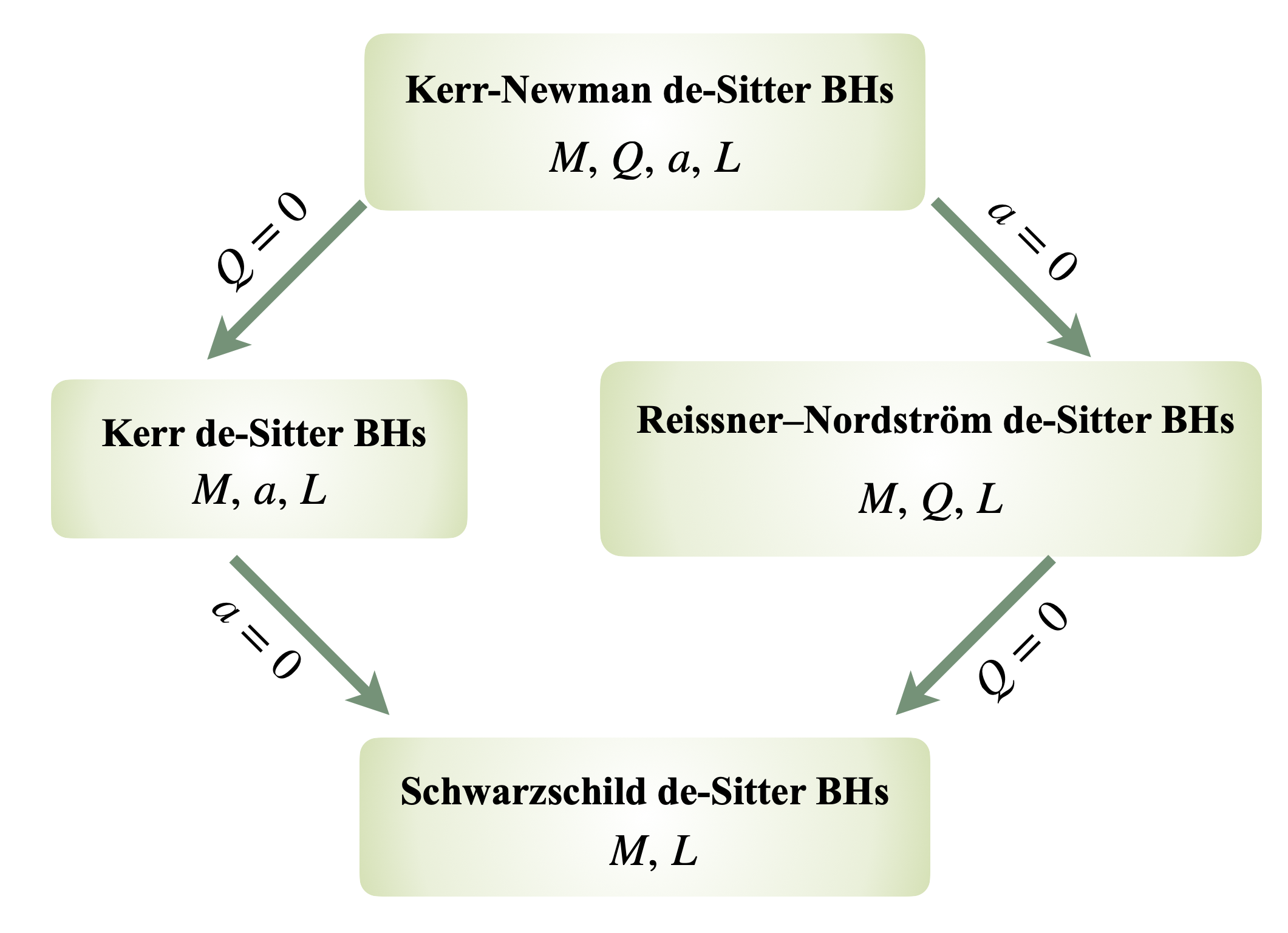}
\caption{The two Nariai limits for rotating charged black holes in de Sitter space with coincident black hole horizon and cosmological horizon: (i) a non-rotating Nariai black hole with $a = 0$, which is the Nariai limit of an RN-dS black hole, and (ii) a rotating Nariari black hole with $Q = 0$, in which the event horizon coincides the cosmological horizon. These Nariai black holes have the common limit of non-rotating and uncharged Nariai black hole, in which the horizon of a Schwarzschild black hole coincides with the cosmological horizon.}
\label{fig_BHrelation}
\end{figure}

The rotating charged Nariai black holes have the geometry of warped ${\rm dS}_3 \times {\rm S}^1/Z_2$ in contrast to the geometry warped ${\rm AdS}_3 \times {\rm S}^1/Z_2$ of near-extremal KN-dS black holes and the geometry ${\rm dS}_2 \times {\rm S}^2$ of (nonrotating) charged Nariai black holes. The angular momentum scaled as $a/L$ with the dS radius $L = \sqrt{3/\Lambda}$ modifies the black hole radius $r_n/L$ and the dS radius $r_\mathrm{ds}$ as well as $M/L$ and $Q/L$. Then, the enhanced symmetry allows one to find the exact solutions of a charged scalar field in terms of the Gauss hypergeometric functions as in the case of charged Nariai black holes with properly modified dimensionless parameters, $Q/L$ and $a/L$. We observe that the mean numbers of spontaneously produced pairs from near-extremal rotating charged Nariai black holes have a universal form in the spacelike region exterior to the cosmological horizon and in the timelike region in-between two horizons. Furthermore, these mean numbers reduce to those in near-extremal (nonrotating) charged Nariai black holes and rotating uncharged Nariai black holes and even to the mean number for nonrotating uncharged Nariai black holes which are the Nariai limit of Schwarzschild-dS black holes, following the flow of diagram in Fig.~\ref{fig_BHrelation}.

The organization of the this paper is as follows. In Sec.~\ref{sec II} we study the rotating charged Nariai black holes and the near-horizon geometry in the timelike inner region between the black hole horizon and the cosmological region and in the spacelike outer region outside of the cosmological horizon. In Sec.~\ref{sec III} we quantize a charged scalar field in both the timelike inner region and the spacelike outer region and define the in-vacuum and out-vacuum states. The mean numbers for the emitted charges via the Schwinger pair production are found in both regions. In Sec.~\ref{sec IV} we discuss the effect of rotation on charge emission and compare the results with those of non-rotating Nariai black holes and of rotating uncharged Nariai black holes.

\section{Rotating Nariai black holes} \label{sec II}
In the Einstein-Maxwell theory with a positive cosmological constant, the Kerr-Newman-de Sitter (KN-dS) solution of a rotating charged black hole is\footnote{In the units of $c = \hbar = 4 \pi \epsilon_0 = 1$, time has the dimension of length, mass the dimension of inverse length, charge dimensionless and $G$ the dimension of length square. Further, setting $G = 1$ makes length, time, mass, and charge all dimensionless. Inserting units, $M \rightarrow G M /c^2$ and $Q^2 \rightarrow G Q^2/4 \pi \epsilon_0 c^4$.}
\begin{eqnarray}
ds^2 &=& - \frac{\Delta_r}{\Sigma} \left( dt - \frac{a \sin^2\theta}{\Xi} d\varphi \right)^2 + \frac{\Sigma}{\Delta_r} dr^2 + \frac{\Sigma}{\Delta_\theta} d\theta^2 + \frac{\Delta_\theta}{\Sigma} \sin^2\theta \left( a dt - \frac{r^2 + a^2}{\Xi} d\varphi \right)^2,
\nonumber\\
A_{[1]} &=& \frac{Q r}{\Sigma} \left( dt - \frac{a \sin^2\theta}{\Xi} d\varphi \right),
\end{eqnarray}
where
\begin{eqnarray}
&& \Delta_r = (r^2 + a^2) \left( 1 - \frac{r^2}{L^2} \right) - 2 M r + Q^2 = - \frac{(r - r_-) (r - r_+) (r - r_c) (r + r_- + r_+ + r_c)}{L^2},
\nonumber\\
&& \Delta_\theta = 1 + \frac{a^2}{L^2} \cos^2\theta, \qquad \Sigma = r^2 + a^2 \cos^2\theta, \qquad \Xi = 1 + \frac{a^2}{L^2}.
\label{param}
\end{eqnarray}
The four parameters (hairs) are the mass $M$, charge $Q$, angular momentum parameter $a = J/M$, and cosmological constant $\Lambda = 3/L^2$ ($L$ being the dS radius). The three horizons, the inner one $r_-$, outer one $r_+$ and cosmological one $r_c$ in increasing order, are related to the dimensionless physical parameters $(M/L, Q/L, a/L)$ as
\begin{eqnarray}
1 - \frac{a^2}{L^2} &=& \frac{1}{L^2}\Bigl( r_+^2 + r_-^2 + r_c^2 + r_+ r_- + r_+ r_c + r_- r_c \Bigr),
\\
\frac{M}{L} + \frac{a^2}{L^2}  \Bigl( \frac{r_+ + r_c}{2L} \Bigr) &=& \frac{r_+ + r_c}{2L}  \Bigl(1  - \frac{r_+^2 + r_c^2}{L^2} \Bigr),
\\
\frac{Q^2}{L^2} + \frac{a^2}{L^2} \Bigl(1 +  \frac{r_+  r_c}{L^2} \Bigr) &=& \frac{r_+ r_c}{L^2} \Bigl( 1 - \frac{ r_+^2 + r_c^2 + r_+ r_c}{L^2} \Bigr).
\end{eqnarray}
The righthand side of equations has the same form as the nonrotating Nariai black hole in dS space except that $r_-, r_+$ and $r_c$ also depend on $a/L$.
The associated Hawking temperature, entropy, horizon angular velocity, and electric potential at the black hole horizon are given by
\begin{eqnarray}
&& T_H = \frac{(r_+ - r_-) (L^2 - a^2 - 3 r_+^2 - 2 r_+ r_- - r_-^2)}{4 \pi (r_+^2 + a^2) L^2} = \frac{(r_c - r_+) (r_c^2 + 2 r_+ r_c + 3 r_+^2 - L^2 + a^2)}{4 \pi (r_+^2 + a^2) L^2},
\nonumber\\
&& S_{BH} = \frac{\pi (r_+^2 + a^2)}{\Xi}, \qquad \Omega_H = \frac{a \Xi}{r_+^2 + a^2}, \qquad \Phi_H = - \frac{Q r_+}{r_+^2 + a^2},
\end{eqnarray}
and the Gibbons-Hawking temperature at the cosmological horizon is
\begin{equation}
T_{GH} = \frac{(r_c - r_+) (r_+^2 + 2 r_+ r_c + 3 r_c^2 - L^2 + a^2)}{4 \pi (r_c^2 + a^2) L^2}.
\end{equation}
The lukewarm limit $T_H = T_{GH}$ leads to the condition $(r_+ + r_c)^2 = L^2 + a^2$, in terms of physical parameters,
\begin{equation}
\frac{M^2}{L^2} = \left( 1 + \frac{a^2}{L^2} \right) \left( \frac{Q^2}{L^2} + \frac{a^2}{L^2} + \frac{a^4}{L^4} \right).
\end{equation}
The limit has a mass bound, $M/L \geq (a/L) (1 + a^2/L^2)$, with the equality for $Q = 0$, and $M = Q$ for nonrotating black holes.

\begin{figure}
\includegraphics[scale=0.45, angle=0]{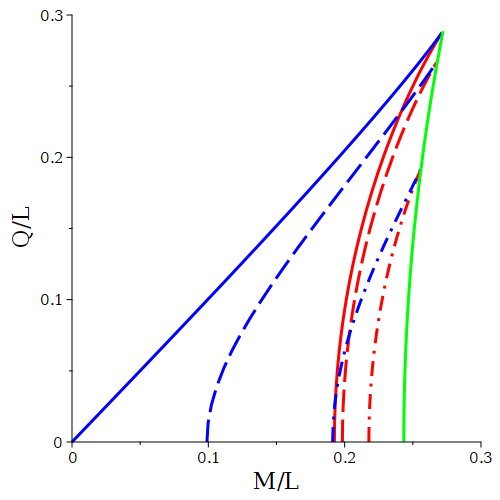}
\caption{Two extremal limits of rotating charged black holes in dS space, the blue curves for the extremal KN black hole with the geometry of a warped $\mathrm{AdS}_3 \times \mathrm{S}^1/Z_2$ and the red curves for the Nariai black hole with the geometry of a warped $\mathrm{dS}_3 \times \mathrm{S}^1/Z_2$, for different angular parameters of $a/L$: (i) the case of $a/L = 0$ (solid lines), (ii) the case of $a/L = 0.1$ (dash lines), and (iii) the case of $a/L = 0.2$ (dash-dot lines). The green solid curve is the ultracold black holes with the three coincident horizons $r_- = r_+ = r_c$, which is given by $Q/L = ( 3^{3/4}/2^{5/6} (M/L)^{4/3} + 2^{7/3} (M/L)^{2/3} - 2^{8/3}/3 )^{1/2}$, and has the range $M/L \in [4(2/\sqrt{3} - 1)^{3/2} = 0.243387, \sqrt{2}/3 \sqrt{3} = 0.272166]$. Note that the ultracold black hole shrinks to a point for nonrotating ones.}
\label{fig_rNariai_QvsM}
\end{figure}

For the charge emission near the black hole horizon and the cosmological horizon, according to Ref.~\cite{Gibbons:1977mu} we use the Killing vector $K = \partial/\partial t$ that becomes timelike, future-directed in the region $r_+ < r < r_c$ and spacelike in the region $r > r_c$. In this paper we will call these regions ``timelike inner region'' and ``spacelike outer region'', respectively.
A KN black hole in dS space can have two coincidence limits for extremal black holes: $r_- = r_+$ or $r_+ = r_c$, where $M, Q, a$ and $L$ satisfy the constraint
\begin{eqnarray}
\left[ \left( 1 - \frac{a^2}{L^2} \right)^3 - 54 \frac{M^2}{L^2} + 36 \left( 1 - \frac{a^2}{L^2} \right) \frac{Q^2 + a^2}{L^2} \right]^2 = \left[ \left( 1 - \frac{a^2}{L^2} \right)^2 - 12 \frac{Q^2 + a^2}{L^2} \right]^3.
\end{eqnarray}
This constraint has two positive real solutions for $M$ corresponding to two possible extremal limits, namely the blue ($r_- = r_+$) or red ($r_+ = r_c$) curves in Fig.~\ref{fig_rNariai_QvsM}. The black holes reach to the ultracold state, i.e. $r_- = r_+ = r_c$, the green line in Fig.~\ref{fig_rNariai_QvsM}, when two solutions of $M$ are degenerated which lead to constraints
\begin{equation}
\frac{Q^2}{L^2} = \frac1{12} \left( \frac{a^4}{L^4} - 14 \frac{a^2}{L^2} + 1 \right), \qquad \frac{M^2}{L^2} = \frac2{27} \left( 1 - \frac{a^2}{L^2} \right)^3.
\end{equation}
For the ultracold state, with a fixed $L$, the $a = 0$ leads to maximum value of $Q$, and similarly $Q = 0$ leads to maximum of $a$
\begin{equation}
\frac{Q_\mathrm{max}}{L} = \frac{1}{2 \sqrt{3}}, \qquad \frac{a_\mathrm{max}}{L} = 2 - \sqrt{3},
\end{equation}
which are two end points of green curve. The Fig.~\ref{fig_rNariai_QvsM} is mathematically the root structure of the algebraic equation $\Delta_r = 0$. The blue ($r_+ = r_-$) and the red ($r_+ = r_c$) curves divide the parameter space into the two regions: the inside region for KN-dS black holes with three distinct roots and the outside region for singular spacetimes with only one root and a naked singularity. It is obvious that (near-)extremal Nariai black holes can evolve to singular spacetimes once the charge emission carries away a significantly large amount of charges. The condition to avoid evolving into singular spacetimes is $q/m < dQ/dM|_{r_+ = r_c}$. The sufficient condition is
\begin{equation} \label{eq_CCScon}
\frac{q}{m} < \frac{dQ}{dM}\Big|_\mathrm{min} = \frac{dQ}{dM}\Big|_{r_- = r_+ = r_c} = \sqrt{\frac{2 (1 - a^2/L^2)}{\bigl( (2 - \sqrt{3})^2 - a^2/L^2 \bigr) \bigl( (2 + \sqrt{3})^2 - a^2/L^2 \bigr)}}.
\end{equation}
In the standard particle model, the sufficient condition can be satisfied when the angular momentum is close to the allowed maximum $a_{\rm max}/L = 2 - \sqrt{3}$, which contrasts to Nariai black holes with the condition $q/m < \sqrt{2}$.

The corresponding physical properties in near-extremal limits have been discussed in~\cite{Anninos:2010gh, Castro:2022cuo}.
The pair production for the near-extremal limit $r_- \sim r_+$ has been discussed in detail~\cite{Chen:2020mqs}. Here we will study the pair production in the other extremal limit, namely Nariai limit as $r_+ = r_c = r_n$ and $M = M_n, Q = Q_n$, where
\begin{equation} \label{eq_NariaiLimit}
r_n^2 = \frac{L^2 - a^2}6 \left( 1 + \sqrt{1 - \frac{12 L^2 (Q_n^2 + a^2)}{(L^2 - a^2)^2}} \right), \qquad M_n = \frac{r_n}3 \left( 1 - \frac{a^2}{L^2} \right) \left( 2 - \sqrt{1 - \frac{12 L^2 (Q_n^2 + a^2)}{(L^2 - a^2)^2}} \right).
\end{equation}
The radius $L$ has a minimal value, $(L_\mathrm{min}^2 - a^2)^2 = 12 L_\mathrm{min}^2 (Q_n^2 + a^2)$, corresponding to the ultracold limit $r_- = r_+ = r_c$. Also note that the angular momentum $a/L$ shifts $r_n/L$ and $M_n/L$ for a nonrotating Nariai black hole by $1 - a^2/L^2$ and $Q_n^2/L^2 + a^2/L^2$. Note that $M_n/L$ increases from the minimum $1/3\sqrt{3}$ as $Q_n/L$ or $a/L$ increases while $r_n/L$ decreases from the maximum $1/\sqrt{3}$ as $Q_n/L$ or $a/L$ increases.

We further consider the near Nariai limit (near-extremal Nariai black hole) with a slight derivation from~\eqref{eq_NariaiLimit} as
\begin{equation} \label{eq_rprc}
r_+ = r_n - \epsilon B, \qquad r_c = r_n + \epsilon B, \qquad M = M_n - \epsilon^2 B^2 \frac{2 r_n}{L^2}, \qquad Q^2 = Q_n^2 - \epsilon^2 B^2 \frac{M_n}{r_n},
\end{equation}
in which $\epsilon \to 0$ but $B$ is fixed.
Then, the Hawking temperature reduces to
\begin{equation}
T_H = \frac{B}{2 \pi} \frac{\epsilon}{r_\mathrm{ds}^2},
\end{equation}
in which an important scale, the radius of dS$_2$ appearing in the near-horizon geometry, is defined as
\begin{equation} \label{eq_rds}
r_\mathrm{ds}^2 = \frac{r_n^2 + a^2}{\Delta_n}, \qquad \Delta_n = \frac{6 r_n^2}{L^2} + \frac{a^2}{L^2} - 1.
\end{equation}
Note that $\Delta_n = (-1/2) d^2 \Delta_r/dr^2 \vert_{r = r_n}$, and the near-horizon geometry has a warped ${\rm dS}_3 \times {\rm S}^1/Z_2$
\begin{equation}
\frac{1}{r_\mathrm{ds}^2} + \frac{1 + 5 a^2/L^2}{r_n^2 + a^2} = \frac{6}{L^2} = 2 \Lambda.
\end{equation}
The near-extremal Nariai has two interesting regions to study the pair production: (i) a spacelike outer region $r > r_c$ and (ii) a timelike inner region $r_+ < r < r_c$, which will be investigated separately below.

\subsection{Spacelike Outer Region}
The geometry of the spacelike outer region of near-extremal Nariai black hole can be represented by a suitable coordinates $(\tau, \rho)$ as
\begin{equation}
\varphi \to \varphi + \frac{a \Xi}{r_n^2 + a^2} \, t, \qquad r = r_n + \epsilon \, \tau, \qquad t = \frac{r_\mathrm{ds}^2}{\epsilon} \, \rho.
\end{equation}
Then by taking $\epsilon \to 0$, one can get the near-horizon geometry ($\tau > B$)
\begin{equation} \label{eq_outNariai}
ds^2 = \frac{\Sigma_n}{\Delta_n} \left( - \frac{d\tau^2}{\tau^2 - B^2} + (\tau^2 - B^2) d\rho^2 + \frac{\Delta_n}{\Delta_\theta} d\theta^2 \right) + \frac{(r_n^2 + a^2)^2 \Delta_\theta \sin^2\theta}{\Sigma_n} \left( \frac{d\varphi}{\Xi} + \frac{2 a r_n \, \tau \, d\rho}{(r_n^2 + a^2) \Delta_n} \right)^2,
\end{equation}
in which the gauge field (the sign of charge is chosen such that the electric field points the positive $\rho$-direction) is given by
\begin{equation}
A_{[1]} = \frac{Q_n (r_n^2 - a^2 \cos^2\theta)}{\Sigma_n \Delta_n} \tau \, d\rho + \frac{Q_n a r_n \sin^2\theta}{\Sigma_n \Xi} d\varphi,
\end{equation}
where
\begin{equation}
\Sigma_n = r_n^2 + a^2 \cos^2\theta.
\end{equation}
The role of $t$ and $r$ is interchanged and~\eqref{eq_outNariai} describes a time-dependent, expanding geometry.

\subsection{Timelike Inner Region}
The near-horizon geometry of $r_+ = r_c = r_n$ can be obtained by applying the following transformation
\begin{equation}
\varphi \to \varphi + \frac{a \Xi}{r_n^2 + a^2} \, t, \qquad r \to r_n + \epsilon \, \rho, \qquad t \to \frac{r_\mathrm{ds}^2}{\epsilon} \, \tau,
\end{equation}
and then taking $\epsilon \to 0$ limit. The parameter $B$ characterizes the deviation from the extremal limit. Finally, the near-horizon geometry of KN-dS black hole ($-B < \rho < B$) is given by
\begin{eqnarray}
ds^2 &=& \frac{\Sigma_n}{\Delta_n} \left( - (B^2 - \rho^2) d\tau^2 + \frac{d\rho^2}{B^2 - \rho^2} + \frac{\Delta_n}{\Delta_\theta} d\theta^2 \right) + \frac{(r_n^2 + a^2)^2 \Delta_\theta \sin^2\theta}{\Sigma_n} \left( \frac{d\varphi}{\Xi} - \frac{2 a r_n \, \rho \, d\tau}{(r_n^2 + a^2) \Delta_n} \right)^2,
\\
A_{[1]} &=& \frac{- Q_n (r_n^2 - a^2 \cos^2\theta)}{\Sigma_n \Delta_n} \rho \, d\tau + \frac{Q_n a r_n \sin^2\theta}{\Sigma_n \Xi} d\varphi.
\end{eqnarray}

\section{Pair Production} \label{sec III}
In this section we study quantum theory for spontaneous emission of charges from the rotating Nariai black hole and thereby the instability of quantum vacuum due to pair production. To do so, we employ the in-out formalism by Schwinger and DeWitt~\cite{DeWitt:2003pm}, which has been elaborated to Schwinger pair production in electromagnetic fields~\cite{Kim:2008yt, Kim:2009pg, Kim:2016nyz} and in (A)dS space~\cite{Cai:2014qba}. In the in-out formalism the Bogoliubov transformations
\begin{eqnarray}
a_{{\rm (out)}\bf K} = \alpha_{\bf K} a_{{\rm (in)}\bf K} + \beta^*_{\bf K} b^{\dagger}_{{\rm (in)}\bf K}, \qquad b_{{\rm (out)}\bf K} = \alpha_{\bf K} b_{{\rm (in)}\bf K} + \beta^*_{\bf K} a^{\dagger}_{{\rm (in)}\bf K},
\end{eqnarray}
for all quantum number ${\bf K}$, relate the out-vacuum to the in-vacuum:
\begin{eqnarray}
a_{{\rm (in)}\bf K} \vert 0, {\rm in} \rangle = b_{{\rm (in)}\bf K} \vert 0, {\rm in} \rangle = 0.
\end{eqnarray}
The Bogoliubov coefficients satisfy the relation from the quantization rule
\begin{equation}
|\alpha_{\bf K}|^2 - |\beta_{\bf K}|^2 = 1.
\end{equation}
The out-vacuum, $a_{{\rm (out)}\bf K} \vert 0, {\rm out} \rangle = b_{{\rm (out)}\bf K} \vert 0, {\rm out} \rangle = 0$, consists of the entangled states
\begin{eqnarray}
\vert 0, {\rm out} \rangle = \prod_{\bf K} \Biggl[ \frac{1}{\alpha_{\bf K}} \sum_{n_{\bf K}} \Bigl( - \frac{\beta^*_{\bf K}}{\alpha_{\bf K}} \Bigr)^{n_{\bf K}} \vert n_{\bf K}, n_{\bf K}, {\rm in} \rangle \Biggr],
\end{eqnarray}
where multi particle-antiparticle states are defined as
\begin{eqnarray}
\vert n_{\bf K}, n_{\bf K}, {\rm in} \rangle = \frac{(a^{\dagger}_{{\rm (in)}\bf K})^{n_{\bf K}}}{\sqrt{n_{\bf K}!}} \frac{(b^{\dagger}_{{\rm (in)}\bf K})^{n_{\bf K}}}{\sqrt{n_{\bf K}!}} \vert 0, {\rm in} \rangle.
\end{eqnarray}
The mean number of spontaneous produced pairs carried by the out-vacuum is
\begin{eqnarray}
{\cal N}_{\bf K} = \langle 0, {\rm out}  \vert  a^{\dagger}_{{\rm (in)}\bf K}  a_{{\rm (in)}\bf K} \vert 0, {\rm out} \rangle = \vert \beta_{\bf K} \vert^2.
\end{eqnarray}

\subsection{Spacelike Outer Region} \label{sec IIIa}
For the outer region, the background spacetime becomes time-dependent, and pair production, from $\tau = B$ to $\tau = \infty$, is analog to a scattering process over a time-dependent potential. Using the ans\"atz
\begin{equation} \label{ansatz}
\Phi_{kln}(\tau, \rho, \theta, \phi) = \mathrm{e}^{i k \rho + i n \phi/\Xi} \, T_{kln}(\tau) S_{ln}(\theta),
\end{equation}
the angular part is the spheroidal function with the separation parameter $\lambda_{l}$
\begin{equation} \label{eq_KGs}
\frac1{\sin\theta} \partial_\theta (\Delta_\theta \sin\theta \partial_\theta S_{ln}) - \left[ \frac{(n \Sigma_n - q Q_n a r_n \sin^2\theta)^2}{(r_n^2 + a^2)^2 \Delta_\theta \sin^2\theta} - m^2 a^2 \sin^2\theta - \lambda_{l} \right] S_{ln} = 0,
\end{equation}
and the radial part is
\begin{equation} \label{eq_KGt}
\partial_\tau \left[ (\tau^2 - B^2) \partial_\tau T_{kln} \right] + \left[ \frac{\left[ q Q_n (r_n^2 - a^2) \tau + 2 n a r_n \tau - (r_n^2 + a^2) \Delta_n k \right]^2}{(r_n^2 + a^2)^2 \Delta_n^2 (\tau^2 - B^2)} + \frac{m^2 (r_n^2 + a^2) + \lambda_{l}}{\Delta_n} \right] T_{kln} = 0.
\end{equation}
The general solution for the KG equation is given by the Gauss hypergeometric function
\begin{eqnarray} \label{sol SOR}
T_{kln}(\tau) &=& c_1 (\tau + B)^{i (\tilde{\kappa} + \kappa)/2} (\tau - B)^{-i (\tilde{\kappa} - \kappa)/2} F\left( \frac12 + i \kappa + i \mu, \frac12 + i \kappa - i \mu; 1 - i \tilde{\kappa} + i \kappa; z \right)
\nonumber\\
&+& c_2 (\tau + B)^{i (\tilde{\kappa} + \kappa)/2} (\tau - B)^{i (\tilde{\kappa} - \kappa)/2} F\left( \frac12 + i \tilde{\kappa} + i \mu, \frac12 + i \tilde{\kappa} - i \mu; 1 + i \tilde{\kappa} - i \kappa; z \right),
\end{eqnarray}
where
\begin{equation}
\tilde{\kappa} = \frac{k}{B}, \qquad \kappa = \frac{q Q_n (r_n^2 - a^2) + 2 n a r_n}{(r_n^2 + a^2) \Delta_n}, \qquad \mu^2 = \kappa^2 + \frac{m^2 (r_n^2 + a^2) + \lambda_{l}}{\Delta_n} - \frac14, \qquad z = - \frac{\tau - B}{2 B}.
\end{equation}
Unlike emissions in the near-extremal black holes ($r_- \simeq r_+$), the pair production threshold $\mu^2 > 0$ does not ensure condition~\eqref{eq_CCScon}, and therefore the singular spacetimes can be physically generated via charge emissions.

\begin{figure}
\includegraphics[scale=0.6, angle=0]{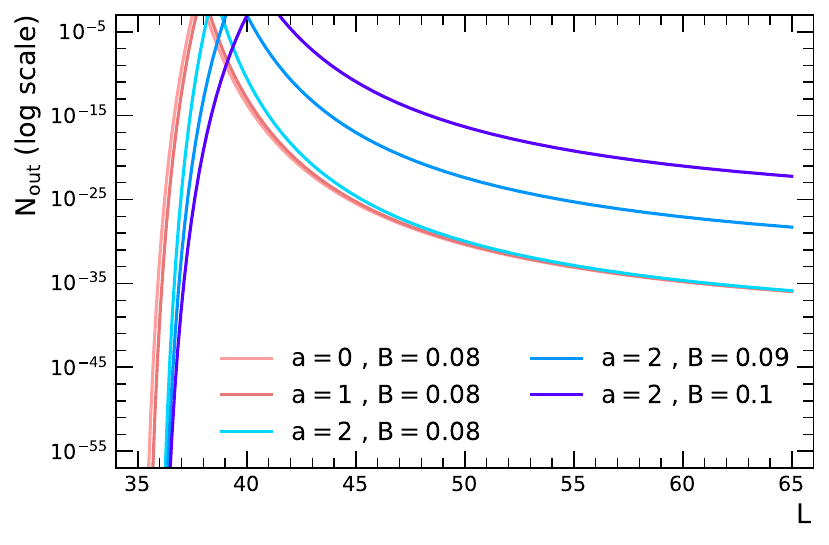}
\hspace{.5cm}
\includegraphics[scale=0.6, angle=0]{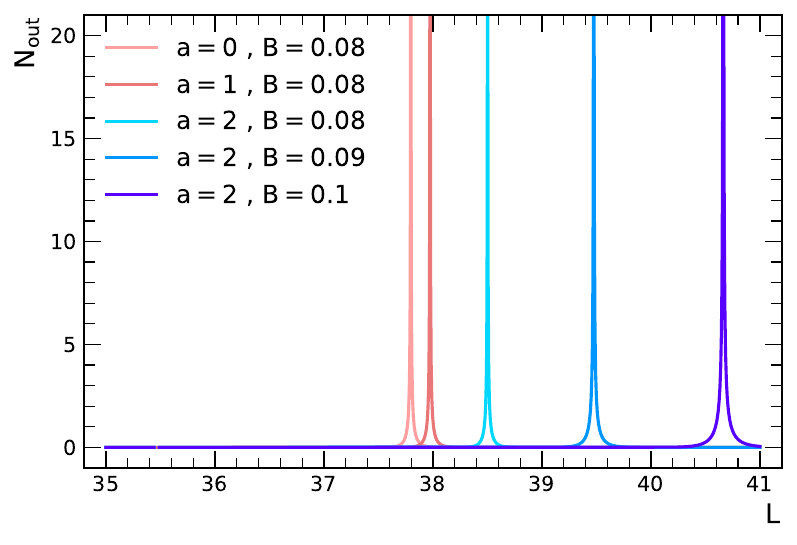}
\caption{[left] The mean number of pairs against $L$ for $a = 0, 1, 2$ with $B = 0.08$ and for $B = 0.08, 0.09, 0.10$ with $a = 2$ in the spacelike outer region. [right] The mean number is amplified for the range $L = [36, 41]$ near the catastrophic emission. Here parameters are fixed at $Q_n = 10, k = 2, \lambda_l = 0, n = 0, m = q = 1$.}
\label{fig_MeanNout}
\end{figure}

The solution~(\ref{sol SOR}) for rotating Nariai black hole has exactly the same form as that for the non-rotating case~\cite{Chen:2023swn} with different values of parameters. So we can easily get the results for the mean number. Note that $k$ is an ``energy'' parameter in the spacelike region.
For the near-extremal limit, $B$ should be small, and $\tilde{\kappa}\, (\ge \kappa)$ is more physical.
\begin{equation} \label{eq_Nout}
\mathcal{N}_\mathrm{out} = \frac{\cosh(\pi \kappa + \pi \mu) \cosh(\pi \tilde{\kappa} - \pi \mu)}{\sinh ( \pi \tilde{\kappa} - \pi \kappa) \sinh(2 \pi \mu)} = \frac{1 + \mathrm{e}^{-2 \pi (\mu + \kappa)}}{1 - \mathrm{e}^{-4 \pi \mu }} \frac{\mathrm{e}^{-2 \pi (\tilde{\kappa} - \kappa)} + \mathrm{e}^{-2 \pi (\mu - \kappa)}}{1 - \mathrm{e}^{-2 \pi (\tilde{\kappa} - \kappa)}}, \qquad \tilde{\kappa} \ge \kappa.
\end{equation}
The case of $\tilde\kappa < \kappa$ is obtained by $\tilde{\kappa} \leftrightarrow \kappa$.\footnote{The mean number for $\tilde\kappa < \kappa$ is $\mathcal{N}_\mathrm{out} = \frac{\cosh(\pi \tilde\kappa + \pi \mu) \cosh(\pi \kappa - \pi \mu)}{\sinh(\pi \kappa - \pi \tilde{\kappa}) \sinh(2 \pi \mu)}$.}
A few comments are in order. First, the form of Eq.~(\ref{eq_Nout}) is a consequence of the scattering boundary condition over the barrier; the $\tau$ is the time variable, and the Coulomb potential becomes timelike. Such an analogue may be observed in the Schwinger pair production in a pulsed Sauter-type electric field $E(t) = E_0/ \cosh^2(t/T)$ with $T$ being an effective duration~\cite{Kim:2008yt}. Second, the most remarkable feature of the mean number is the catastrophic emission of a boson condensation near $\tilde\kappa \approx \kappa$. The physical meaning of the catastrophic emission will be manifest in terms of the effective temperature and Hawking temperature below. Third, as shown in Fig.~\ref{fig_MeanNout}, for a fixed $B$, increasing the angular momentum of black hole shifts the peak of the mean number toward further large dS radius $L$ while for a fixed angular momentum, increasing $B$ moves further the peak for large dS radius.
The angular momentum of black hole decreases the mean number.

We interpret the mean number in terms of the effective temperature. We write the mean number as
\begin{equation} \label{eq_Nout2}
\mathcal{N}_\mathrm{out} = \mathrm{e}^{-2 \pi(\mu - \kappa)} \frac{1 + \mathrm{e}^{-2 \pi (\mu + \kappa)}}{1 - \mathrm{e}^{-2 \pi (\mu - \kappa)} \mathrm{e}^{-2 \pi (\mu + \kappa)}} \frac{1 + \mathrm{e}^{-2 \pi (\tilde{\kappa} - \kappa)} \mathrm{e}^{2 \pi (\mu - \kappa)}}{1 - \mathrm{e}^{-2 \pi (\tilde{\kappa} - \kappa)}}, \qquad \tilde{\kappa} \ge \kappa.
\end{equation}
By introducing an ``effective inertial mass'' $\bar{m}$ as in Ref.~\cite{Chen:2023swn},
\begin{equation}
\mu^2 - \kappa^2 = m^2 r_\mathrm{ds}^2 + \frac{\lambda_l}{\Delta_n} - \frac14 = \bar{m}^2 r_\mathrm{ds}^2,
\end{equation}
where $r_\mathrm{ds}$ is defined in~\eqref{eq_rds}, the corresponding thermodynamic quantities can be obtained, whose detailed discussions can be found in~\cite{Chen:2023swn}, from the following relations
\begin{eqnarray}
2 \pi (\mu + \kappa) &=& \frac{\mu^2 - \kappa^2}{(\mu - \kappa)/2 \pi} = \frac{\bar{m}^2 r_\mathrm{ds}^2}{\left( \sqrt{\kappa^2 + \bar{m}^2 r_\mathrm{ds}^2} - \kappa \right)/2 \pi} = \frac{\bar m}{\sqrt{T_U^2 + T_C^2} - T_U},
\nonumber\\
2 \pi (\mu - \kappa) &=& \frac{\mu^2 - \kappa^2}{(\mu + \kappa)/2 \pi} = \frac{\bar{m}^2 r_\mathrm{ds}^2}{\left( \sqrt{\kappa^2 + \bar{m}^2 r_\mathrm{ds}^2} + \kappa \right)/2 \pi} = \frac{\bar m}{\sqrt{T_U^2 + T_C^2} + T_U},
\\
2 \pi (\tilde{\kappa} - \kappa) &=& 2 \pi \frac{k}{B} - 2 \pi \frac{q Q_n (r_n^2 - a^2) + 2 n a r_n}{(r_n^2 + a^2) \Delta_n} = \frac{k - q \Phi_H - n \Omega_H}{T_H},
\nonumber
\end{eqnarray}
where $T_U$ is the Unruh temperature, $T_C$ is the temperature associated to the dS curvature, and $T_H, \Phi_H, \Omega_H$ are the Hawking temperature (in rescaled coordinates), chemical (electric) potential, and horizon angular velocity, respectively,
\begin{equation} \label{temperatures}
T_U = \frac{\kappa}{2 \pi \bar{m} r_\mathrm{ds}^2}, \qquad T_C = \frac1{2 \pi r_\mathrm{ds}}, \qquad T_H = \frac{B}{2 \pi}, \qquad \Phi_H = \frac{Q_n (r_n^2 - a^2) B}{(r_n^2 + a^2) \Delta_n}, \qquad \Omega_H = \frac{2 a r_n B}{(r_n^2 + a^2) \Delta_n}.
\end{equation}

Finally, we can express the mean number in terms of the effective temperatures as
\begin{equation} \label{mean-num-space}
\mathcal{N}_\mathrm{out} = \mathrm{e}^{-\bar{m}/T_\mathrm{eff}} \frac{1 + \mathrm{e}^{-\bar{m}/\bar{T}_\mathrm{eff}}}{1 - \mathrm{e}^{-\bar{m}/T_\mathrm{eff}} \mathrm{e}^{-\bar{m}/{\bar T}_\mathrm{eff}}} \frac{1 + \mathrm{e}^{-(k - q \Phi_H - n \Omega_H)/T_H} \mathrm{e}^{\bar{m}/T_\mathrm{eff}}}{1 - \mathrm{e}^{-(k - q \Phi_H - n \Omega_H)/T_H} },
\end{equation}
where
\begin{equation} \label{eff-tem}
T_\mathrm{eff} = \sqrt{T_U^2 + T_C^2} + T_U, \qquad {\bar T}_\mathrm{eff} = \sqrt{T_U^2 + T_C^2} - T_U.
\end{equation}
A few comments again are in order. First, the mean number exhibits a universal formula as in~\cite{Chen:2023swn}. Second, the dS radius $r_\mathrm{ds}$ increases as $a/L$ or $Q/L$ increases, so the rotation reduces the Gibbons-Hawking temperature $T_C$. Third, the rotation decreases both the Unruh temperature $T_U$ and the effective temperature $T_\mathrm{eff}$ for charges for with $n = 0$. Fourth, the Boltzmann factor $\mathrm{e}^{-\bar{m}/T_\mathrm{eff}}$ is the leading term because $T_\mathrm{eff} \gg \bar{T}_\mathrm{eff}$. Finally, the catastrophic emission occurs when $| (k - n \Omega_H) - q \Phi_H | \ll T_H$; that is, the effective energy, $k - n \Omega_H$, differs from the chemical potential, $q \Phi_H$, by small amount in units of the Hawking temperature.
Note that the mean number can be written as
\begin{equation}
\mathcal{N}_\mathrm{out} = \frac{1 + \mathrm{e}^{-\bar{m}/\bar{T}_\mathrm{eff}}}{\mathrm{e}^{\bar{m}/T_\mathrm{eff}}- \mathrm{e}^{-\bar{m}/{\bar T}_\mathrm{eff}}} \frac{\mathrm{e}^{(k - q \Phi_H - n \Omega_H)/T_H} + \mathrm{e}^{\bar{m}/T_\mathrm{eff}}}{\mathrm{e}^{(k - q \Phi_H - n \Omega_H)/T_H} - 1}.
\end{equation}
The first factor is the mean number for the Schwinger effect in ${\rm dS}_2$ and the second factor is the Hawking radiation for a near-extremal black hole. The numerators are amplification factors or greybody factors, respectively.

For rotating Nariai black hole, we take $B = 0 \; (T_H = 0)$ limit and obtain
\begin{eqnarray} \label{Nariai-emission}
\mathcal{N}_\mathrm{out} = \frac{\mathrm{e}^{-2 \pi (\mu - \kappa)} + \mathrm{e}^{-4 \pi \mu} }{1 - \mathrm{e}^{-4 \pi \mu}}
= \frac{\mathrm{e}^{-\bar{m}/T_\mathrm{eff}} + \mathrm{e}^{-\bar{m}/[ T_C /( 2 \sqrt{1 + T_U^2/T_C^2} ) ]}}{1 - \mathrm{e}^{-\bar{m}/[ T_C /( 2 \sqrt{1 + T_U^2/T_C^2} ) ]}}.
\end{eqnarray}
The emission formula~\eqref{Nariai-emission} recovers the (nonrotating) Nariai black holes and also rotating uncharged Nariai black holes.

\subsection{Timelike Inner Region} \label{sec IIIb}
In the timelike inner region, the background spacetime is static, and pair production, from $\rho = -B$ to $\rho = B$, is analog to a tunneling process through a potential barrier. The scalar field is decomposed into the spherical harmonic and a positive frequency
\begin{equation} \label{ansatzInt}
\Phi_{\omega ln}(\tau, \rho, \theta, \phi) = \mathrm{e}^{-i \omega \tau + i n \phi/\Xi} \, R_{\omega ln}(\rho) S_{ln}(\theta),
\end{equation}
the angular equation is just~\eqref{eq_KGs} and then the radial mode of the KG equation for $R(\rho)$ reduces to
\begin{equation}
\partial_\rho \left[ (B^2 - \rho^2) \partial_\rho R_{\omega ln} \right] + \left[ \frac{\left[ q Q_n (r_n^2 - a^2) \rho + 2 n a r_n \rho - (r_n^2 + a^2) \Delta_n \omega \right]^2}{(r_n^2 + a^2)^2 \Delta_n^2 (B^2 - \rho^2)} - \frac{m^2 (r_n^2 + a^2) + \lambda_l}{\Delta_n} \right] R_{\omega ln} = 0.
\end{equation}
The general solution of the KG equation is given by the Gauss hypergeometric function
\begin{eqnarray}
R_{\omega ln}(\rho) &=& c_1 (B + \rho)^{-i (\tilde{\kappa} + \kappa)/2} (B - \rho)^{-i (\tilde{\kappa} - \kappa)/2} F\left( \frac12 - i \tilde{\kappa} + i \mu, \frac12 - i \tilde{\kappa} - i \mu; 1 - i \tilde{\kappa} + i \kappa; z \right)
\nonumber\\
&+& c_2 (B + \rho)^{-i (\tilde{\kappa} + \kappa)/2} (B - \rho)^{i (\tilde{\kappa} - \kappa)/2} F\left( \frac12 - i \kappa + i \mu, \frac12 - i \kappa - i \mu; 1 + i \tilde{\kappa} - i \kappa; z \right),
\end{eqnarray}
where
\begin{equation} \label{t-parameters}
\tilde{\kappa} = \frac{\omega}{B}, \qquad \kappa = \frac{q Q_n (r_n^2 - a^2) + 2 n a r_n}{(r_n^2 + a^2) \Delta_n}, \qquad \mu^2 = \kappa^2 + \frac{m^2 (r_n^2 + a^2) + \lambda_l}{\Delta_n} - \frac14, \qquad z = - \frac{\rho - B}{2 B}.
\end{equation}

The problem describes a tunneling process and the mean number of pair production is\footnote{The mean number for $\tilde{\kappa} < \kappa$ again can be obtained from~\eqref{eq_Nin} by exchanging $\tilde{\kappa}$ and $\kappa$.}
\begin{equation} \label{eq_Nin}
\mathcal{N}_\mathrm{in} = \frac{\sinh(\pi \tilde{\kappa} + \pi \kappa) \sinh(\pi \tilde{\kappa} - \pi \kappa)}{\cosh(\pi \kappa + \pi \mu) \cosh(\pi \kappa - \pi \mu)} = \mathrm{e}^{2 \pi (\tilde{\kappa} - \kappa)} \mathrm{e}^{-2 \pi(\mu - \kappa)} \frac{(1 - \mathrm{e}^{-2 \pi (\tilde{\kappa} + \kappa)}) (1 - \mathrm{e}^{-2 \pi (\tilde{\kappa} - \kappa)})}{(1 + \mathrm{e}^{-2 \pi (\mu + \kappa)}) (1 + \mathrm{e}^{-2 \pi (\mu - \kappa)})}, \qquad \tilde{\kappa} \ge \kappa.
\end{equation}
Note that the form of Eq.~(\ref{eq_Nin}) is a consequence of the tunneling boundary condition. An analogue may be found in the Schwinger pair production in a spatially localized Sauter-type electric field $E(z) = E_0 / \cosh^2(z/L)$ with $L$ being an effective width~\cite{Kim:2009pg}. The amplification of emission has the form $\mathrm{e}^{2 \pi (\tilde{\kappa} - \mu)}$, in contrast to the boson condensation-like catastrophic emission in the spacelike outer region.

In the limit $B \to 0$, i.e. $\tilde{\kappa} \to \infty$, the tunneling region shrinks to a spherical surface of zero volume and thus the barrier disappears, making the tunneling ``trivial.'' Consequently the mean number diverges and the charge emission becomes catastrophic
\begin{equation}
\lim_{\tilde{\kappa} \to \infty} \mathcal{N}_\mathrm{in} = \mathrm{e}^{2 \pi \tilde{\kappa}} \to \infty.
\end{equation}
In fact, the thermal interpretation for the mean number shows a universal behavior as in \cite{Chen:2023swn}
\begin{equation} \label{exp-amplification}
\mathcal{N}_\mathrm{in} = \mathrm{e}^{(\omega - q \Phi_H - n \Omega_H)/T_H} \mathrm{e}^{- \bar{m}/T_\mathrm{eff}} \frac{\left( 1 - \mathrm{e}^{-(\omega + q \Phi_H + n \Omega_H)/T_H} \right) \left( 1 - \mathrm{e}^{-(\omega - q \Phi_H - n \Omega_H)/T_H} \right)}{\left( 1 + \mathrm{e}^{-\bar{m}/T_\mathrm{eff}} \right) \left( 1 + \mathrm{e}^{-\bar{m}/\bar{T}_\mathrm{eff}} \right)},
\end{equation}
and implies that charges with energy larger than chemical potential are exponentially produced $\mathrm{e}^{(\omega - q \Phi_H - n \Omega_H)/T_H}$ and charge emission exponentially explodes as the distance between two horizons draws closer and closer.

\begin{figure}
\includegraphics[scale=0.6, angle=0]{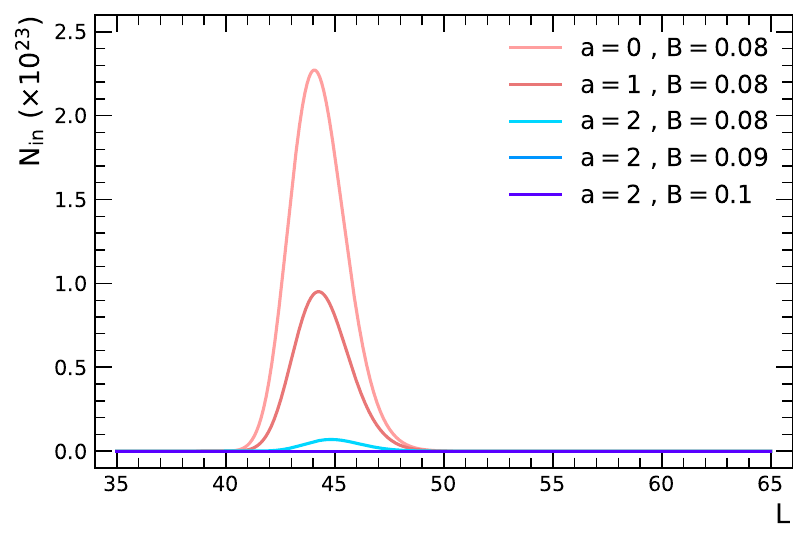}
\hspace{.5cm}
\includegraphics[scale=0.6, angle=0]{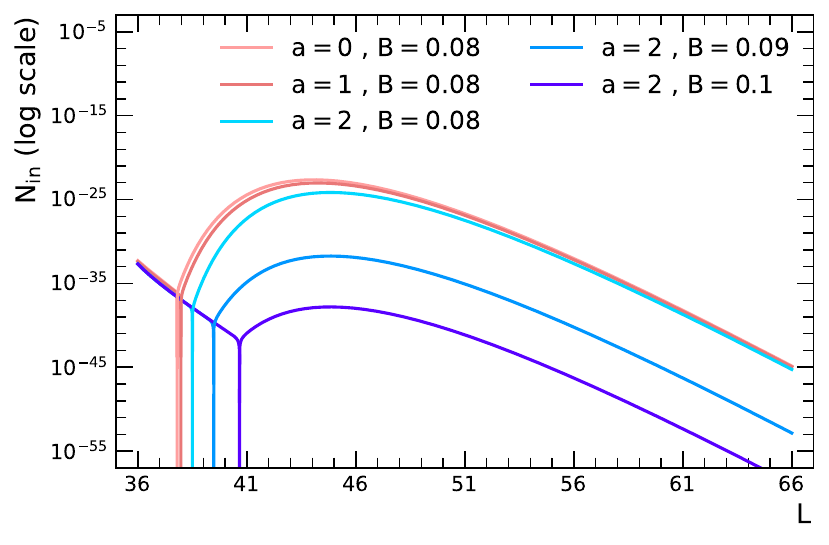}
\caption{The mean number of pairs against $L$ in the timelike inner region: [left] $\mathcal{N}_\mathrm{in}$ for $a = 0, 1, 2$ with $B = 0.08$ and for $B = 0.08, 0.09, 0.10$ with $a = 2$ and [right] the mean number in the log scale. Here parameters are fixed at $Q_n = 10, \omega = 2, \lambda_l = 0, n = 0, m = q = 1$.}
\label{fig_MeanNin}
\end{figure}

\subsection{Effects of Angular Momentum and Azimuthal Number} \label{sec IIIc}
In this subsection, we study the effects of angular momentum parameter $a$ and azimuthal number $n$ on the mean number $\mathcal{N}$. In subsections~\ref{sec IIIa} and~\ref{sec IIIb} we have confined the mean number for the emission to charges with $n = 0$. However, any non-zero azimuthal quantum number $n$ also contributes to the Unruh temperature through $\kappa$, which contrasts to (non-rotating) charged Nariai black holes only with $\lambda_l = l(l+1)$ but without dependence on the azimuthal quantum number~\cite{Chen:2023swn}.

In fact, the effective mass for a charge with mass $m$ is given by
\begin{eqnarray}
\bar{m}^2 = m^2 + \frac{\lambda_l}{r_n^2} - \frac{1}{4 r_\mathrm{ds}^2},
\end{eqnarray}
which reduces to that of the same charge with $\lambda_l = l(l+1)$ in the case of (non-rotating) charged Nariai black holes. On the other hand, the azimuthal quantum number $n \neq 0$ contributes an additional term, the second term, to $\kappa$
\begin{eqnarray}
\kappa (a, n) = \kappa (a, 0) + \frac{2n a r_n}{(r_n^2 + a^2) \Delta_n}.
\end{eqnarray}
The Gibbons-Hawking temperature $T_C = 1/(2 \pi r_\mathrm{ds})$ is independent of $n$ but the Unruh temperature depends on $n$ as
\begin{eqnarray}
T_U (a, n) = T_U (a, 0) + n \Biggl(\frac{a r_n}{\pi \bar{m} r_\mathrm{ds}^2 (r_n^2 + a^2) \Delta_n} \Biggr).
\end{eqnarray}
Note that as $a$ increases, $T_C$ always decreases regardless of $n$ while $T_U$ monotonically increases ($dT_U/da > 0$ for $n \neq 0$), which make $T_\mathrm{eff}$ a concave function of $a$ ($dT_\mathrm{eff}/da$ changes signs about the peak).

\begin{figure}
\includegraphics[scale=0.6, angle=0]{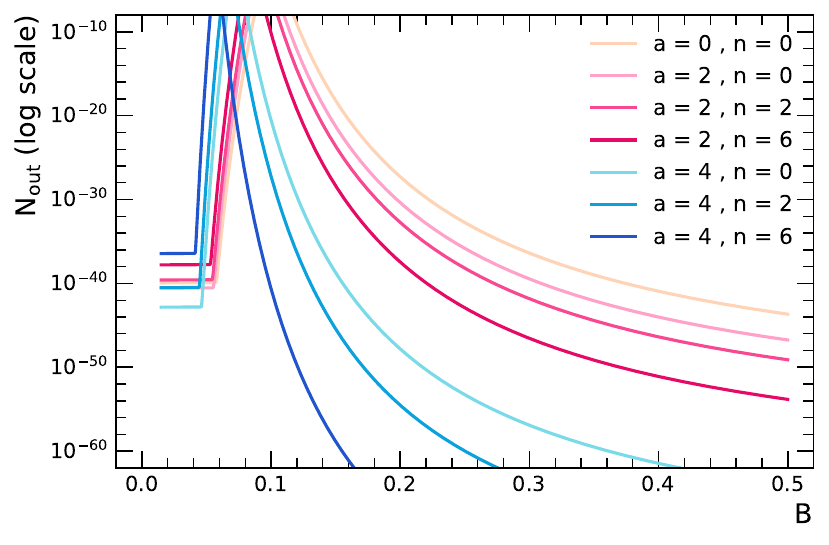}
\hspace{.5cm}
\includegraphics[scale=0.6, angle=0]{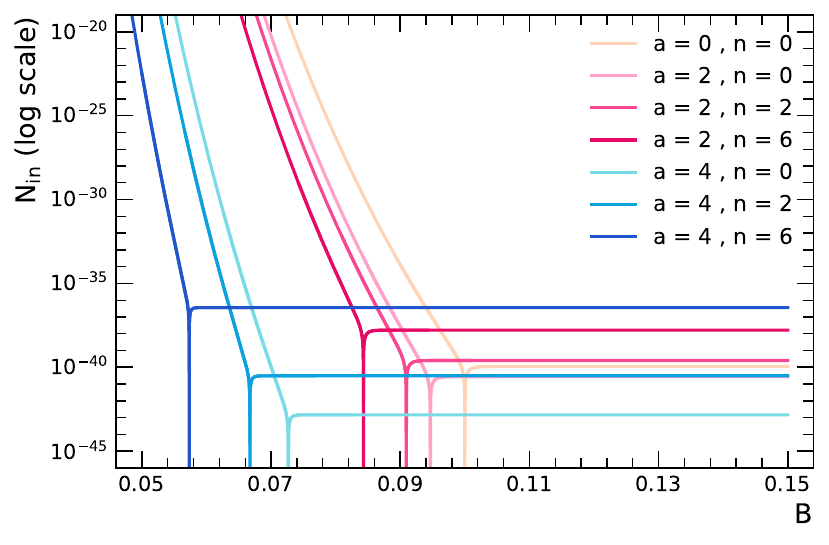}
\caption{The mean number of pairs against $B$ in the spacelike outer region [left] and timelike inner region [right]. Here parameters are fixed at $L = 40, Q_n = 10, \lambda_l = 0, m = q = 1$, and $k = 2$ (outer) or $\omega = 2$ (inner). The parameter $a$ is chosen in the range $a/L \le 0.1 < 2 - \sqrt3 = a_\mathrm{max}/L$.}
\label{fig_N_vs_a}
\end{figure}

The mean numbers with respect to parameter $B$ are presented in Fig.~\ref{fig_N_vs_a}, $\mathcal{N}_\mathrm{out}$ in the spacelike outer region on the left, and $\mathcal{N}_\mathrm{in}$ in the timelike inner region on the right, with various choices of angular momentum per unit mass $a$ and azimuthal number $n$. The parameter $B$ basically represents a tiny separation between the outer horizon $r_+$ and cosmological horizon $r_c$, as defined in~\eqref{eq_rprc}, for near extremal Nariai black holes. As the results in Figs.~\ref{fig_MeanNout} and ~\ref{fig_MeanNin} show, the mean number has a special limit, $\mathcal{N}_\mathrm{out}$ diverges and $\mathcal{N}_\mathrm{in}$ vanishes, at $\tilde\omega = \omega$. For the consideration of near extremal limit, the results of small $B$, i.e. $\tilde\omega > \omega$ are more essential.\footnote{The effects $a$ and $n$ for $\tilde\omega < \omega$ are actually opposite.} In the inner region, the effects of both $a$ and $n$ reduce the mean number leading to suppress the catastrophic emission, namely occurring at a smaller value of $B$. Consequently, the catastrophic emission ceases if $a$ is larger than the critical value $a_\mathrm{max}/L = 2 - \sqrt{3}$. However, in the outer region with $\tilde\omega < \omega$, contrary to the inner region, the mean number is enhanced by both $a$ and $n$, and behaves like $\mathcal{N}_\mathrm{out} \propto \tanh\tilde\omega$ as $B \to 0$ (i.e. $\tilde\omega \to \infty$). The ``flat'' tails in the figure are log-scale enhanced ``turning'' of function $\tanh(x)$ around $x = 1$.

\subsection{Comparison with Other Extremal Black Holes} \label{sec IIId}
Now we compare the emission formula for rotating charged Nariai black holes with those for charged Nariai black holes~\cite{Chen:2023swn} and near-extremal KN-dS black holes~\cite{Chen:2021jwy}. First, charged Nariai black holes have the near-horizon geometry of $\mathrm{dS}_2 \times \mathrm{S}^2$~\cite{Ortaggio:2002bp}:
\begin{eqnarray}
\frac{1}{r_\mathrm{ds}^2} + \frac{1}{r_n^2} = \frac{6}{L^2},
\end{eqnarray}
where
\begin{eqnarray}
r_n^2 = \frac{L^2}6 \left( 1 + \sqrt{1 - \frac{12 Q_n^2}{L^2}} \right), \qquad \Delta_n = \frac{6 r_n^2}{L^2} - 1, \qquad r_\mathrm{ds}^2 = \frac{r_n^2}{\Delta_n}.
\end{eqnarray}
It contrasts with the geometry warped $\mathrm{dS}_3 \times \mathrm{S}^1/Z_2$ of rotating charged Nariai black holes. The parameters $\kappa$ and $\mu$ for charged Nariai black holes are the $a = 0$ limit of those in Eq.~(\ref{t-parameters}) with $\lambda_l = l (l + 1)$, and similarly the effective mass and temperatures are the same limit of $a = 0$. The results have been discussed in our previous paper~\cite{Chen:2023swn}.

Second, the near-extremal KN-dS black holes have the near-horizon geometry of warped $\mathrm{AdS}_3 \times \mathrm{S}^1/Z_2$:
\begin{eqnarray}
- \frac{1}{r_\mathrm{ads}^2} + \frac{1 + 5 a^2/L^2}{r_0^2 + a^2} = \frac{6}{L^2},
\end{eqnarray}
where $r_0$ is the black hole radius and
\begin{eqnarray}
r_\mathrm{ads}^2 = \frac{r_0^2 + a^2}{\Delta_0}, \qquad \Delta_0 = 1 - \frac{a^2}{L^2} - \frac{6 r_0^2}{L^2}.
\end{eqnarray}
Note that $\Delta_0 = - \Delta_n|_{r_n \to r_0}$ and that $r_\mathrm{ds} \geq r_\mathrm{ads}$ for all allowed values of $a/L$ and $Q/L$ and the equality for the ultracold black holes.
The effective mass felt by charges on the horizon
\begin{eqnarray}
\bar{m}_\mathrm{ads}^2 = m^2 + \frac{\lambda_l + \Delta_0/4}{r_{\rm ads}^2 \Delta_0},
\end{eqnarray}
is larger than that of rotating charged Nariai black holes. Moreover, the effective temperature
\begin{eqnarray}
T_\mathrm{eff} = T_U + \sqrt{T_U^2 \mp \Bigl( \frac{1}{2 \pi r_\mathrm{(a)ds}} \Bigr)^2},
\end{eqnarray}
exhibits a big difference that extremal KN-dS black holes have the upper sign while rotating charged Nariai black holes have the lower sign. The minus sign for extremal KN-dS black holes gives the BF bound $1/(2 \pi r_\mathrm{ads}) > T_U$ for stability against charge emission. Further, rotating charged Nariai black holes have higher effective temperature and emit charges more efficiently than near-extremal KN-dS black holes, not to mention an exponential amplification factor~(\ref{exp-amplification}).

Third, we consider the charge neutral limit, $Q = 0$. Kerr-dS black holes still have the Nariai limit, in which the black hole horizon coincides with the cosmological horizon. All the results of Secs.~II and~III are valid for $Q = 0$ limit. Particularly, note that
\begin{eqnarray}
r_n^2 = \frac16 \left( L^2 - a^2 + \sqrt{L^4 - 14 L^2 a^2 + a^4} \right), \qquad \Delta_n = \frac{6 r_n^2}{L^2} - 1, \qquad  r_\mathrm{ds}^2 = \frac{r_n^2 + a^2}{\Delta_n},
\end{eqnarray}
and thus
\begin{eqnarray}
\kappa = \frac{2 n a r_n}{(r_n^2 + a^2) \Delta_n}, \qquad T_U = \frac{\kappa}{2 \pi \bar{m} r_\mathrm{ds}^2 }, \qquad T_C = \frac{1}{2 \pi r_\mathrm{ds}}, \qquad T_B = \frac{B}{2 \pi}.
\end{eqnarray}
Except for the zero azimuthal quantum number $(n = 0)$ or zero angular momentum $(a = 0)$ that leads to $T_U = 0$ and $\Omega_H = 0$, dragging of charges gives $T_U \neq 0$, needless to say $T_C \neq 0$ and $T_B \neq 0$. Further, the rotational chemical potential is proportional to $\kappa$ as $\Omega_H = \kappa B$.

Finally, the limit of $Q = 0$ and $a = 0$, that is, Schwarzschild-dS black holes also have the Nariai limit, whose emission~(\ref{eq_Nout}) in the spacelike outer region describes spherical radiations of massive particles:
\begin{eqnarray}
{\cal N}_\mathrm{out} = \frac{1}{\mathrm{e}^{2 \pi \mu} -1} \times \frac{1 + \mathrm{e}^{-2 \pi (\tilde{\kappa} - \mu)}}{1 - \mathrm{e}^{- 2 \pi \tilde{\kappa}}}, \qquad \mu = \sqrt{m^2 r_\mathrm{ds}^2 + l (l + 1) - \frac{1}{4}},
\end{eqnarray}
where the first factor is the Gibbons-Hawking radiation with the temperature $T_C = 1/(2 \pi r_\mathrm{ds})$ in dS space and $\tilde{\kappa} = k/B \gg 1$. In the timelike inner region the emission (\ref{eq_Nin}) reads ${\cal N}_\mathrm{in} = (\sinh \pi \tilde{\kappa}/\cosh \pi \mu)^2$ with $\tilde{\kappa} = \omega/B \gg 1$. In the limit $B \to 0$, the ${\cal N}_\mathrm{in}$ becomes catastrophic and ${\cal N}_\mathrm{out}$ reduces to pure de Sitter.

In conclusion, the Schwinger emission formulas~(\ref{mean-num-space}) and~(\ref{exp-amplification}) for near-extremal rotating charged Nariai black holes exhibit a universality and are the most general in that following Fig.~\ref{fig_BHrelation} they reduce to the formulas for near-extremal charged Nariai black holes, near-extremal rotating uncharged Nariai black holes and then near-extremal nonrotating uncharged Nariai black holes, which are the Nariai limit of Schwarzschild-dS black holes.

\section{Conclusion} \label{sec IV}

Rotating charged (Kerr-Newman) black holes in de Sitter space have three horizons: the inner (Cauchy) horizon, the black hole (outer) horizon, and the cosmological horizon. The coincidence limit of the inner horizon and black hole horizon gives the (near-)extremal Kerr-Newman black hole. On the other hand, the coincidence limit of the black hole horizon and the cosmological horizon gives rotating charged Nariai black holes. Extremal black holes have zero Hawking temperature and a mere extrapolation of Hawking radiation in a literal sense still seems to lead to radiation with effective energy equal to the chemical potential. However, the (near-)extremal black holes have a different near-horizon geometry from the Rinder space for non-extremal ones, and the Hawking radiation does not apply to the extremal black holes. The enhanced symmetry of near-horizon of (near-)extremal black holes allows one to solve the field equation for charges and find the emission formula due to the so-called Schwinger mechanism~\cite{Chen:2012zn, Chen:2014yfa, Chen:2016caa, Chen:2017mnm, Chen:2020mqs, Chen:2021jwy}.

The (near-)extremal Kerr-Newman black holes and Narial black holes exhibit big differences. The most noticeable difference is the Breitenlohner-Freedman (BF) bound for extremal Kerr-Newman black holes, which results from the near-horizon geometry of warped ${\rm AdS}_3 \times {\rm S}^1/Z_2$~\cite{Chen:2021jwy}. On the other hand, the near-horizon geometry of rotating charged Nariai black holes is a warped ${\rm dS}_3 \times {\rm S}^1/Z_2$, which does not have the BF bound and continuously emits charges via the Schwinger mechanism in the dS space. In fact, the dS temperature increases the effective temperature for accelerating charges in the electric field on black hole horizon. In the previous paper, we have shown that non-rotating charged Nariai black holes have a catastrophic emission of charges, a kind of boson condensation, in the spacelike outer region of cosmological horizon and an exponential amplification for emission for charges with high energy~\cite{Chen:2023swn}. A noticeable aspect is that the rotation effect can rescue the (nonrotating) charged Nariai black holes from evolving into singular spacetimes with a naked singularity by emitting Schwinger pairs.

In this paper, we have studied the near-horizon geometry of rotating charged Nariai black holes in which the black hole horizon approaches close to the cosmological horizon and the quantum field theory for a charged scalar field to find an explicit formula for the charge emission. Particularly, we have focused on the effect of rotation on the emission of charges. The emission formulae for charged Nariai black holes exhibit a universal factorization that reflects the near-horizon geometry of (near-)extremal black holes. We have shown that the near-horizon geometry of rotating charged Nariai black holes, a warped ${\rm dS}_3 \times {\rm S}^1/Z_2$, can catastrophically enhance the emission of charges partly because the dS-temperature contributes to the effective temperature without the BF bound as for (nonrotating) charged Nariai black holes, and more importantly because there is an exponential amplification factor in the timelike inner region between the two horizons or a boson condensation in the spacetime outer region of the cosmological horizon.

On the other hand, the rotation of Nariai black holes suppresses the emission of charges by a factor not by an order because the black hole radius increases as the angular momentum increases, and thus the reduced electric field in the horizon lowers the effective temperature. A physically interesting observation is that rotating charged Nariai black holes, unlike the nonrotating case, can avoid evolving into singular spacetimes (a naked singularity) through charge emission. However, a complete evolution of charged Nariai black holes must include the back-reaction of emitted charges, which goes beyond of this work and will be addressed in a future work. Finally, we have compared the emission formula of charges from rotating charged Nariai black holes with those of charged Nariai black holes and (near-)extremal Kerr-Newman black holes in de Sitter space.

\acknowledgments
The work of C.M.C. was supported by the National Science and Technology Council of the R.O.C. (Taiwan) under the grant NSTC 113-2112-M-008-027.
The work of S.P.K. was supported in part by National Research Foundation of Korea (NRF) funded by the Ministry of Education (2019R1I1A3A01063183).


\end{document}